# Profitable Strategy Design by Using Deep Reinforcement Learning for Trades on Cryptocurrency Markets

Mohsen Asgari[#1], Seyed Hossein Khasteh[#2]

[#]*Artificial Intelligence Department, Faculty of Computer Engineering, K. N. Toosi University of Technology*

[1] Mohsen0Asgari@gmail.com
[2] khasteh@kntu.ac.ir

*Abstract*— Deep Reinforcement Learning solutions have been applied to different control problems with outperforming and promising results. In this research work we have applied Proximal Policy Optimization, Soft Actor-Critic and Generative Adversarial Imitation Learning to strategy design problem of three cryptocurrency markets. Our input data includes price data and technical indicators. We have implemented a Gym environment based on cryptocurrency markets to be used with the algorithms. Our test results on unseen data shows a great potential for this approach in helping investors with an expert system to exploit the market and gain profit. Our highest gain for an unseen 66 day span is 4850$ per 10000$ investment. We also discuss on how a specific hyperparameter in the environment design can be used to adjust risk in the generated strategies.

## I. INTRODUCTION

Since the birthplace of the term "Artificial Intelligence" in scientific literature at a workshop at Dartmouth College in 1956 by John McCarthy (McCorduck, 2004) (Crevier, 1993) till today this field of science has dramatically changed our perspectives to many problems. Nowadays with a wide range of different approaches and algorithms, this discipline thrives to shape our future even more. Market direction prediction is one of the most challenging and also one of the most rewarding subfields of applications of AI in finance. A good prediction for the markets direction can help investors to allocate their capital on assets where they can expect to obtain the most promising rewards. If a tool can provide an effective insight about this process, it can play a crucial rule for individual investors and enterprise firms for their decisions on markets and effecting their gain and loss in their investment.

Investment is the propulsion force in an economy. Good decisions in choosing the best and the most profitable sections of a market not only can provide benefits for the investor but also it will participate in growth and development of the sections with higher potential in providing goods and services to end users, hence making investment decisions one of the most critical parts of policies for market and economy participants.

Nowadays, a very important aspect of most cryptocurrencies is their independence from governments and pretty much any financial institutions. This attribution of these assets makes them a perfect fit for pure markets dynamic analysis because external factors could be excluded from them and also makes it easy for individuals and independent firms to participate in the market. This ease in participation in a global market itself creates an attractiveness for development of new and innovative approaches for the analysis due to increment of the potential customers and users of these services. Not to forget these users usually hold a competitive intentions against each other's strategy.

Using experts in financial decision making is not a new thing. Experts in mathematical and statistical methods have been in action since 1900 and there is even a job title for those who work in this area: Quantitative Analysts or for short "Quants". These experts whom usually are very expensive to hire have developed different frameworks for different aspects of financial decision making. But AI has casted a shadow on this field like many other fields. A lot of studies have shown outperformance of artificial intelligence methods in comparison with conventional approaches. A great survey done on this topic has been provided by (Bahrammirzaee, 2010).

We have discussed some of important works done on this topic in the upcoming section. Main challenge of AI and ML based methods in the topic of financial decision making is to produce a steady flow of returns and avoidance of losses in different market conditions. Other challenges include how the proposed system deals with risk management issues and how much complex system would be, regarding the data and process it needs.

Traditionally, one of the main machine learning approaches has been known as Reinforcement Learning (RL), where unlike supervised learning there is no need to provide the machine with example inputs and their desired outputs. Instead, RL algorithms interact with an environment and receive a feedback signal from it. Usually the paradigm is set to maximize this feedback signal (we can alternatively call it the reward signal).

This approach has fundamental similarities with how animal's (and human's) psychology works for example by interpreting hunger and pain signals as negative reinforcements and pleasure and food as positive signals (Daeyeol Lee, 2012; Stuart J. Russell, 2010, pp. 830-831).

This agent-based approach is suitable for times when we don't know the best action in our environment but we can measure our agent's success in achieving its goal by a feedback signal. This looks very fit for strategy creation for profitable trades on financial markets.

In our modelling of the strategy creating problem, we have a simulated market where it has been built using historical data obtained from that market. The agent observes current timeframe data and also 59 data points before it and it uses three different deep reinforcement learning algorithms (namely, Proximal Policy Optimization (PPO) (F. W. John Schulman, Prafulla Dhariwal, Alec Radford, Oleg Klimov, 2017), Soft Actor Critic (SAC) (Tuomas Haarnoja, 2018) and Generative Adversarial Imitation Learning (GAIL) (Jonathan Ho, 2016)) to devise a policy for doing profitable trades.

In our previous work (Asgari & Khasteh, 2021) we have studied using three statistical learning approaches for strategy creation in cryptocurrency markets. In this research work we focus on three different state-of-the-art deep reinforcement learning methods to model our problem and craft solutions for it.

The main goal of the methods described in this article are to drive a policy for deciding when and how much should the agent buy or sell the cryptocurrencies in order to achieve profit. As a metaphor, this looks like a trader tries to learn market dynamics by analysing previous historical movements of the price and then uses obtained knowledge to make profitable decisions in the market.

Training data for this system has been gathered by using a data channel directed to the Binance cryptocurrency exchange API. Then we run some preprocessing procedures on the data and get them ready to be used as entry to our machine learning models.

Three different deep reinforcement learning algorithms have been used in this work. All of them categorise as Policy Gradient methods. We have discussed in detail these models in the "Methods and Materials" section.

Data used for these analyses are mostly Open, High, Low, Close and Volume data from three different cryptocurrency markets: ETH-USDT, LTC-BTC, ZEC-BTC. We have used 4 hour timeframe for our experiments. These data have been augmented by technical indicators to make better learning data for models.

After explaining the models and the data, we explore the implementation of these algorithms and our modelling of the market in the "Proposed Method" section.

At the "Experimental Results" section we look at the performance of these models in the test data which our learned models have not been exposed to.

At the "Discussion" section we look at the performance of our models and discuss some different and debatable aspects of these methods and the whole strategy creation system. We specifically point to some hyperparameters which has a substantial effect on the performance and risk managements in this setting. There are some improvements which can be made to this work and we mention some of them in the "Conclusion and Future Works" section.

## II. RELATED WORKS

First part of this section introduces some previous attempts in using artificial intelligence methods in decision making on financial markets and points to their findings and conclusions. It has been organized based on the publication date of the research works.

After that we look at three different surveys done on the topic of using deep reinforcement learning on financial markets and also point to the methods used in this research work and their previous applications in other studies.

Some researchers denote the emergence of study and application of artificial intelligence in finance back to 1991 by works of Abramson and Finizza on oil markets (Zuzana, 2021). Their work consisted of creating an artificial intelligence knowledge base to model all variables which they thought could have an impact on oil market (Abramson & Finizza, 1991). Kim and Chun in 1998 have evaluated several backpropagation models to predict stock market index where it made an influential effect on future studies (Kim & Chun, 1998). In 1999 Quah and Srinivasan in (Quah & Srinivasan, 1999) said due to ANN's generalizing ability they are able to derive the characteristics of executive stocks from historical patterns. In 2004 Se-Hak and Kim in (Se-Hak & Kim, 2004) showed that the error for forecasting markets by Case Base Reasoning (CBR), which is an artificial intelligence technique, was lower than using neural networks and both were better than random walk. Their works also showed that it is better to generate several intermediate points for forecasting and interpolate with them than generating a prediction for end point at one step. In 2007 Oj and Kim showed Case Base Reasoning can be used in prediction of collapse of a financial market (Oj & Kim, 2007).

Since 2009 number of publications in this field increases. This coincides with the mortgage crisis which was the biggest financial crisis with global impact since last 30 years. At this time rate of publications per year in Web of Science networks doubles from previous rate (the rate between 2000 until 2009) (Zuzana, 2021). It can be inferred that after this event, attention of researchers in using AI in finance has been increased dramatically.

Work of Hadavandi et al. in 2010 (Hadavandi, Shavandi, & Ghanbari, 2010) by combining a genetic fuzzy system with ANNs for creating an expert system for stock price prediction is one of most cited works in after the crisis period. Their results showed that the proposed method is outperforming all previous methods hence can be considered a suitable tool for stock price prediction.

From 2010 until 2020 a lot of advances have been happened in artificial intelligence methods and processing powers. The impact of these improvements can be also seen in financial applications too. In this period more attention has been given to combined artificial intelligence methods and hybrid models (Zuzana, 2021). In 2015 Vella & Ng looked at application of AI in control risk-based money management decisions. They constructed a fuzzy logic approach for identification and categorization of technical rules. Their system was proposed for intraday level and it worked by dynamically prioritizing better performing regions in the market and adapting money management methods to maximize global risk-adjusted returns. Their model computes the approximate risk-adjusted performance at each step and then autonomously balances the money in decided allocations. This study found a hybrid method using fuzzy rules combined with a NN for trend prediction works better than standard NN and buy and hold approach (Vella & Ng, 2015).

Works of Qui et al. in 2016 (Qiu & Akagi, 2016) showed that using selective input variables can enhance predictive power of machine learning methods for returns in stock markets. Works of Wei in 2016 with the aim of improving the performance of time series prediction proposed a hybrid model of Adaptive Network Fuzzy Inference system (ANFIS) which relied on Empirical Decomposition Model (EMD) (Wei, 2016).

In 2020 we see another topic arising in the literature which can show the impact of AI in finance to be even more momentous. Lee in his works (Lee, 2020) discusses legal and regulatory necessities for usage of AI in financial services markets. Their works includes a proposal for a framework for these regulations. They also present policy recommendations and some gaudiness for usage of AI in finance.

Another interesting study in 2020 is from Arrieda et el. (Arrieta, et al., 2020) on their literature on "Explainable AI" where they point to "Black-Box" problem of some of AI methods which is in contrast with "Responsible AI" concept where large-scale implementation of AI methods in real organizations has been obliged with fairness, model explainability and accountability at their core.

Ruan et el. In (Ruan, Wang, Zhou, & Lv, 2020) proposed the idea that investors sentiments can have a heavy impact on assets prices based on the assumption that many investors are not thoroughly relied on their rationale when trading in the market and psychological factors affect them. They used deep neural networks to design an indicator which was based on investor's sentiment. Their predictions using this technique has outperformed other widely recognized predictors and showing that AI can be very effective in estimating sentiments.

Both long-term and short-term decisions have been point of interest for researchers in this field, but Millner & Heyen in (Millner & Heyen, 2021) state that short-term decision making systems are of higher priority as they can be more useful for mangers to alter their decisions effectively. Also Millner & Heyen point to more difficult nature of long-term predictions due to the fact that they require new scientific approaches that reduce model misspecification in comparison with short-term predictions where they often can be substantially improved by simply reducing measurement errors in initial conditions or in other words by increasing the quality of observations. Petrelli and Cesarini in (Petrelli & Cesarini, 2021) specifically recommend usage of AI in high frequency trading systems.

Many researchers suggest that all businesses, especially the ones facing a competitive relationship with each other, will have to apply and adopt to AI approaches because failure to deployment of these methods can put their business in danger when competed with firms which they use AI (Milana & Ashta, 2021). It is worthy to mention that AI application goes beyond analytics of data in this field as it can provide a feedback loop and allow itself to learn from its own trial and interactions with the environment. For the short terms, AI-based decision-making is likely to perform better where decisions are very specific. More general decisions may still require a human factor.

To sum up this part we present some visualized bibliometric analysis done by (Zuzana, 2021) about publications in the field of AI in finance. Figure 1 shows how the number of publications and citations for individual years have been developing since 30 years ago. Figure 2 shows a keyword occurrence analysis in the related papers. Figure 3 shows an analysis about co-citations between the journals publishing the research works. All the papers used for these analysis have been gathered from Web of Science Network between 1991 till 2020.

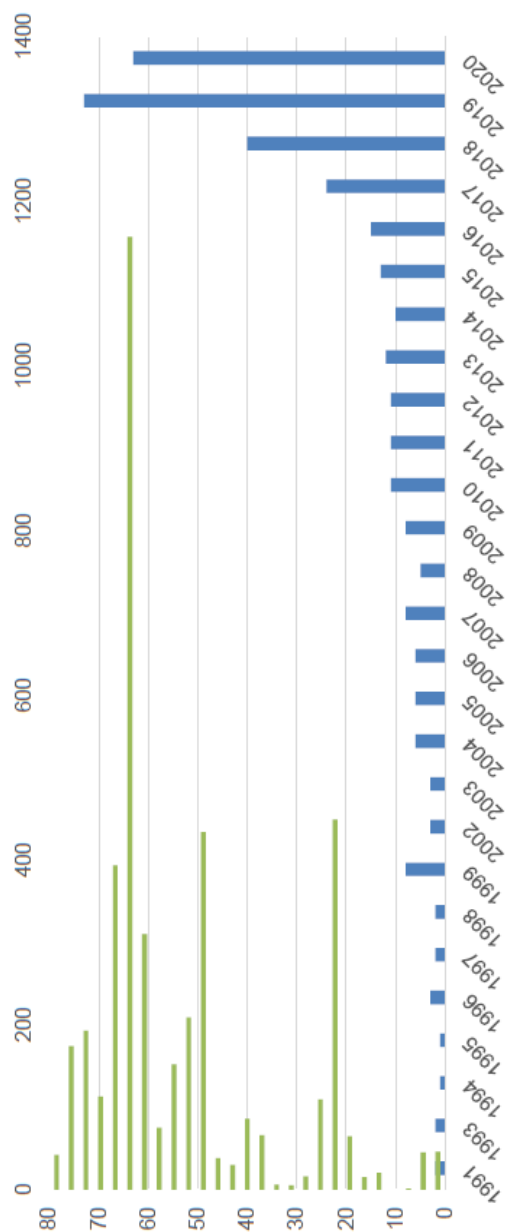

Figure 1. Development of the number of publications and citations for individual years from (Zuzana, 2021)

Figure 2. Keyword occurrence analysis from (Zuzana, 2021)

Figure 3. Analysis of journal co-citations from (Zuzana, 2021)

In the survey's part, first survey (Fischer, 2018) discusses deep reinforcement learning algorithms in financial markets in three categories: Critic-Only Approaches, Actor-Only Approaches and Actor-Critic Approaches. Regarding advantages of RL algorithms for financial markets, it points that "RL allows to combine the "prediction" and the "portfolio construction" task in one integrated step, thereby closely aligning the machine learning problem with the objectives of the investor". This survey has conducted a comprehensive view (including details about Key Findings, States Representation, Actions and Reward Function Designs) of different attempts for using RL algorithms in this context. Based on the findings of this survey, "critic-only approach is the most popular RL approach in financial markets". With mentioning (J. Moody, 2001) it also suggests: "actor-only approach currently appears to be the best suited approach for financial markets" due to availability of continuous actions, small number of parameters, good convergence behaviour and also recent improvements with deep learning techniques.

Second survey (Sato, 2019) suggests if we define our portfolio which consists of multiple investment components as the system being controlled and component weights as the controllers, the problem could be solved using model-free Reinforcement Learning without knowing specific component dynamics. It notes "The biggest disadvantage of the valued-based RL methods (such as Q-Learning) is Bellman's curse of dimensionality that arises from large state and action spaces, making it difficult for the agent to efficiently explore large action spaces". For policy-based methods it points that "approximating the optimal policy with a neural network with many parameters (including hyperparameters) is difficult and can suffer from suboptimal solutions mainly due to its instability, sample inefficiency, and sensitivity on the selection of hyperparameter values". This survey also points to two interesting issues about research works done in this area. One of them is state space is underrepresented where we assume the past asset return is a good predictor of the future asset return. It suggests more "meaningful" features, e.g., fundamentals data or market sentiment data to be used in the state representation of the market. The second issue is the interpretability of machine learning models (Weijs, 2018), since institutional investors do not want to risk a large amount of capital in a model that cannot be explained by financial or economic theories, nor in a model for which the human portfolio manager cannot be responsible. Almost all of the "deep" RL methods use a neural network and neural network is infamously known as "black box" in the discipline. So, this may affect the operability of these methods.

Third survey (Mosavi, et al., 2020) has categorized its studied methods in Reinforcement Learning, Deep Learning and Deep Reinforcement Learning groups. Beside using these approaches for Financial Market Prediction and Portfolio Allocation, it includes use cases in other economic challenges e.g. Business Intelligence, Unemployment Prediction and Risk Management. It notes "exponential increase" in popularity of deep reinforcement learning applications in economics. It points to DRL's potential in high-dimensional problems in conjunction with noisy and nonlinear patterns of economic data and also its better performance and higher efficiency as compared to the traditional algorithms while facing real economic problems in the presence of risk parameters and the ever-increasing uncertainties. This research work has suggested employment of a variety of both DL and deep RL techniques, with the relevant strengths and weaknesses, to serve in economic problems to enable the machine to detect the optimal strategy associated with the market.

Proximal Policy Optimization is a new family of policy gradient methods for reinforcement learning introduced by (Schulman, Wolski, Dhariwal, Radford, & Klimov, 2017) which has been designed to keep benefits of Trust Region Policy Optimization (TRPO) (Schulman, Levine, Abbeel, Jordan, & Moritz, 2015) while being much simpler to implement, more general and having better sample complexity. Experiment tests on a collection of benchmark tasks, including simulated robotic locomotion and Atari game playing, has shown that PPO outperforms other online policy gradient methods (Schulman, Wolski, Dhariwal, Radford, & Klimov, 2017). A study on using this algorithm on 14 US equities has been report to achieve mean gain-loss ratio of 1.23 between October 2018 and December 2018 (Lin & Beling, 2020). Another study on Chinese stock market has reported annual return of higher than 55% with using PPO on their test set (Yuan, Wen, & Yang, 2020).

Soft Actor-Critic (SAC) has been introduced by (Haarnoja, Zhou, Abbeel, & Levine, 2018) as an off-policy RL algorithm to tackle high sample complexity and brittleness to hyperparameters. A systematically evaluation of SAC on a range of benchmark tasks has been reported to achieves state-of-the-art performance, outperforming prior on-policy and off-policy methods in sample-efficiency and asymptotic performance (Haarnoja, Zhou, Abbeel, & Levine, 2018). A study of this approach on Chinese Stock Market has reported higher than 30% annual return (Yuan, Wen, & Yang, 2020).

Generative Adversarial Imitation Learning has been designed by (Ho & Ermon, Generative adversarial imitation learning, 2016) in order to imitate complex behaviours in large, high dimensional environments. This method has been designed to overcome speed issues in using inverse reinforcement learning approaches to learning a policy from example expert behaviour, without interacting with the expert or access to a reinforcement signal. This approach can be used to derive a policy from observed expert behaviours in financial markets, hence reverse engineering the knowledge and methods of the expert. To best of our knowledge this method has not yet been studied in the literature as a solution to financial strategy creation.

### III. METHODS AND MATERIALS

In this section we first look at the data used in this project, then we get acquainted with three different methods which have been used to make the models for the strategy creation task.

#### 1. Used Data

Binance is a cryptocurrency exchange that provides a platform for trading various cryptocurrencies. As of April 2021, Binance was the largest cryptocurrency exchange in the world in terms of trading volume (Top Cryptocurrency Spot Exchanges, 2021).

Binance provides a free to use API for data gathering. This API is conveniently available to use in python language. We use this API to gather Time stamp (in second precision), Open, High, Low, Close and Volume for a 4 hours period dataframe. This procedure runs for all three different assets that we study: ETH-USDT, LTC-BTC and ZEC-BTC. Data gets gathered from mid-2017 until April 2021. This makes our raw input data.

#### 2. First RL Algorithm: Proximal Policy Optimization

Policy Gradient methods are a family of RL algorithms where we don't drive our policy from a value function, instead the policy itself gets improved over time. These methods work by computing an estimator of the policy gradient and plugging it into a stochastic gradient *ascent* algorithm (Note that we want to maximise the policy output values, hence we want to use its gradient in ascending the function). Main challenges in dealing with policy gradient methods are their sensitivity to the choice of stepsize and their very poor sample efficiency.

The main objective in PPO algorithm is the following:

$$L^{CLIP}(\theta) = \hat{E}_t[\min(r_t(\theta)\hat{A}_t, clip(r_t(\theta), 1 - \epsilon, 1 + \epsilon)\hat{A}_t)]$$

where $\pi_\theta$ is a stochastic policy, $\hat{A}_t$ is an estimator of the advantage function at timestep $t$, $\epsilon$ is a hyperparameter and $r_t(\theta)$ denotes the probability ratio

$$r_t(\theta) = \frac{\pi_\theta(a_t|s_t)}{\pi_{\theta_{Old}}(a_t|s_t)}$$

The motivation behind this formula is to optimize the objective function while ensuring the deviation from the previous policy is relatively small.

Proximal policy optimization methods have the stability and reliability of trust-region methods (Schulman, Levine, Abbeel, Jordan, & Moritz, 2015) but are much simpler to implement, requiring only few lines of code change to a vanilla policy gradient implementation, applicable in more general settings (for example, when using a joint architecture for the policy and value function), and have better overall performance (Schulman, Wolski, Dhariwal, Radford, & Klimov, 2017).

#### 3. Second RL Algorithm: Soft Actor-Critic

As mentioned in the previous subsection a great challenge in using recent RL algorithms such as TRPO, PPO and Asynchronous Actor-Critic Agents

(A3C) (Mnih, et al., 2016) is their suffrage from sample inefficiency. This is because they learn in an "on-policy" manner, i.e. they need completely new samples after each policy update. In contrast, Q-learning based "off-policy" methods such as Deep Deterministic Policy Gradient (DDPG) (Lillicrap, et al., 2015) and Twin Delayed Deep Deterministic Policy Gradient (TD3PG) (Dankwa & Zheng, 2019) are able to learn efficiently from past samples using experience replay buffers. However, the problem with these methods is that they are very sensitive to hyperparameters and require a lot of tuning to get them to converge properly. Soft Actor-Critic is an off-policy method which tries to confront the mentioned issue with modifying RL objective function in a way that also addresses the exploration-exploitation dilemma.

Main objective function of SAC is

$$J(\pi) = \sum_{t=0}^{T} E_{(s_t,a_t) \sim \rho_\pi} [r(s_t, a_t) + \alpha \mathcal{H}(\pi(.|s_t))]$$

where $\sum_t E_{(s_t,a_t) \sim \rho_\pi}[r(s_t,a_t)]$ denotes the expected sum of rewards and $\mathcal{H}(\pi(.|s_t))$ denotes the entropy term. We can think of entropy as how unpredictable a random variable is. If a random variable always takes a single value then it has zero entropy because it's not unpredictable at all. If a random variable can be any Real Number with equal probability then it has very high entropy as it is very unpredictable. SAC tries to get a high entropy in the policy to explicitly encourage exploration, to encourage the policy to assign equal probabilities to actions that have same or nearly equal Q-values, and also to ensure that it does not collapse into repeatedly selecting a particular action that could exploit some inconsistency in the approximated Q function. Therefore, SAC overcomes the brittleness problem by encouraging the policy network to explore and not assign a very high probability to any one part of the range of actions. (Kumar, 2019) So SAC has better sample efficiency than PPO and also it tunes entropy coefficient by putting the entropy term inside the objective function.

4. *Third RL Algorithm: Generative Adversarial Imitation Learning*

RL methods need well-defined reward functions, which tell the agents how well they are doing. Unfortunately, in a majority of real-world situations, the reward functions are usually hard to define.

Inverse Reinforcement Learning (IRL) was introduced to help RL learning experts' policy and get reward functions to explain the experts' behaviors from the given experts' trajectories. However, for most classical IRL methods, a large number of expert trajectories should be provided, but in many cases, the trajectories are not easy to get.

Imitation learning techniques "aim to mimic human behaviour in a given task" (Hussein, Gaber, Elyan, & Jayne, 2017). Learning by imitation facilitates teaching complex tasks with minimal expert knowledge of the tasks. Generic imitation learning methods could reduce the problem of teaching a task to provide demonstrations without the need for explicit programming or designing reward functions specific to the task (Gonfalonieri, 2021). As mentioned in various research papers, generative adversarial imitation learning demonstrates "tremendous success in a limited number of use cases, especially when combined with neural network parameterization" (Zhang, Cai, Yang, & Wang, 2020). Generative Adversarial Imitation Learning (GAIL) could be defined as a model-free imitation learning algorithm. This algorithm has shown impressive performance gains compared with other model-free methods in imitating complex behaviours, especially in large, high-dimensional environments.

GAIL algorithm works based on Generative Adversarial Networks (GANs) (Goodfellow, et al., 2014). In a GAN, two networks exist: the generator and the discriminator. The generator's role is to generate new data points by learning the distribution of the input data. The discriminator's part is to classify whether a given data point is generated by the generator (learned distribution) or real data distribution. Thanks to the combination of IRL and GAN concepts, GAIL can learn from a small number of expert trajectories. Indeed, GAIL's goal is to train generators that have similar behaviours to the given experts. In the meantime, the discriminators can serve as the reward functions for RL, which judge whether the behaviours look like the experts. In GAIL, the discriminator learns to distinguish generated performances from expert demonstrations. Simultaneously, the generator attempts to mimic the expert to fool the discriminator into thinking its performance was an expert demonstration.

In GAIL algorithm, goal is to find a saddle point $(\pi, D)$ of the expression

$$E_\pi[\log(D(s,a))] + E_{\pi_E}[\log(1 - D(s,a))] - \lambda H(\pi)$$

To do so, algorithm designers have introduced function approximation for $\pi$ and $D$: algorithm will fit a parameterized policy $\pi_\theta$, with weights $\theta$, and a discriminator network $D_w: S \times A \rightarrow (0,1)$, with weights w. Then, it alternates between an Adam (Kingma & Ba, 2014) gradient step on $w$ to increase the above expression with respect to $D$, and a TRPO step on $\theta$ to decrese the expression with respect to $\pi$. The TRPO step serves the same purpose as it does with the apprenticeship learning algorithm in (Ho, Gupta, & Ermon, Model-free imitation learning with policy optimization, 2016): it prevents the policy from changing too much due to noise in the policy gradient. The discriminator network can be interpreted as a local cost function providing learning signal to the policy—specifically, taking a policy step that decreases expected cost with respect to the cost function $c(s,a) = \log D(s,a)$ will move toward expert-like regions of state-action space, as classified by the discriminator (Ho & Ermon, Generative adversarial imitation learning, 2016).

## IV. PROPOSED METHODS

In this section we look at our proposed methods for strategy creation for cryptocurrency markets using DRL algorithms. First, we fill in details about our raw data gathering procedure. Then at the second subsection, we elaborate on our pre-processing steps for the obtained raw financial data. Third subsection explains our environment designed for these RL tasks and also our state representation, possible actions and reward function. The forth step, sums up the definition of our three different models. We also make some concise points about hyper parameters of each model. Last subsection looks at evaluation of results and concludes the strategy creation part of the system. Figure 5 (look at page 12) shows a comprehensive view of the whole system, green lines indicate train phase and red lines indicate exertion phase.

### 1. Raw Data Gathering

At the time of doing this research, Binance has made available, access to its historical records (for Open, High, Low, Close and Volume) through its API for time frames bigger than one minute. We gather 4 hour period data to a Pandas dataframe since its first available timestamp (which is usually mid 2017). The data includes OHLCV for ETH-USDT, LTC-BTC, ZEC-BTC. We use 95% of the data for training and the remaining 5% to evaluate the models.

### 2. Pre Processing

After gathering the data, we augment it with some famous Technical Indicators in finance. Typically these indicators are some mathematical functions which take some arguments from real time or past data and they create some insights about "technical" moves of the market. Name and formula for these technical indicators has been reported in the Appendix A. We also normalize the data by removing the mean and scaling to unit variance for each feature. After augmentation we have a dataframe including all relevant data for each record.

### 3. Environment Setup

We have inspired our environment design by works of (Liu, et al., 2020). The designed environment is an implemented Gym Environment Class. Gym (Brockman, et al., 2016) is an open source Python library for developing and comparing reinforcement learning algorithms by providing a standard API to communicate between learning algorithms and environments, as well as a standard set of environments compliant with that API. We have also used the open source implementation of algorithms in (Hill, et al., 2018).

State representation in this system includes 1082 continues features. These features include account balance of the agent at the moment, asset balance of the agent at the moment, normalized values of closed price and other augmented financial data through past 60 timeframes and also previous return of the asset (difference between current value of the asset and its immediate predecessor value). Figure 4 shows a schematic view of our designed state space.

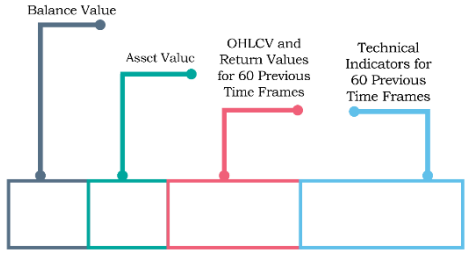

Figure 4. A Schematic View of The State Representation in The Environment Setup

Action space in this setup is a floating point variable between -1 and +1. As shown in the below equation, to get the Sell/Buy Amount, action value gets multiplied by a hyperparameter called Maximum Buying Amount. This hyperparameter is directly related to risk management abilities of this trading paradigm. We have discussed it more in "Discussion" section of this work.

$$Sell\ or\ Buy\ Amount = Maximum\ Buying\ Amount * Action\ Value$$

If the agent tries to buy or sell more than its available balances, environment takes the action at the maximum possible amount. To facilitate the learning process we have added some penalty to the agent denoted with "$\xi$" (as part of its reward signal) each time it exceeds its available balances.

We define "step" as each time our agent has made the observation and took the action and produced new states. Per each step in running algorithm we calculate the difference between "the asset value plus account balance value" (The Gross Value of the account) before and after it and multiply that amount to a reward scalar to normalize it. This scalar factor has amount of $10^{-4}$ for our cryptocurrency pairs. Resulting value ($\xi$) makes the main part of the reward signal and sums with previously introduced value to make the final reward signal at each step. Final reward signal's formula is the following:

$$Gross\ Value_t = Asset\ Value_t + Balance\ Value_t$$
$$Reward\ Signal = [(Gross\ Value_t - Gross\ Value_{t-1}) * v] + \xi$$

4. *Model definition- Model Training*

At this subsection we look at the hyperparameters involved in each model in the implementation that we have used based on OpenAI Gym and Stable Baselines. We also note each model's training time.

A more elaborate discussion about the hyper parameters is held in the Discussion section.

Hyper parameters involved in Proximal Policy Optimization algorithm are as follows:

*gamma (Discount factor):* 0.99

*n_steps (The number of steps to run for each environment per update):* 32

*ent_coef (Entropy coefficient for the loss calculation)*: 0.005

*learning_rate: 0.005*

*used architecture for policy representation:* Multi Layer Perceptron - 2 layers of 64

*total_timesteps (The total number of samples to train on):* 200000

Hyper parameters involved in Soft Actor Critic algorithm are as follows:

*gamma (Discount factor):* 0.99

*learning_rate:* 0.01

*buffer_size (Size of the replay buffer):* 1000

*batch_size (Minibatch size for each gradient update):* 1000

*ent_coef* (*Entropy regularization coefficient):* auto (using 0.1 as initial value)

*learning_starts* (*How many steps of the model to collect transitions for before learning starts*): 200

*used architecture for policy representation:* Multi Layer Perceptron - 2 layers of 64

*total_timesteps* (*The total number of samples to train on):* 50000

An important note about using GAIL algorithm is that it needs some expert behaviours to be trained on. For this research work we have used the policy derived from PPO algorithm (with the above hyper parameters) to generate these behaviours, but the best results with using this algorithm probably will be achieved if we feed it with real successful traders actions.

Other hyper parameters involved in Generative Adversarial Imitation Learning algorithm are as follows:

*gamma (Discount factor):* 0.99

*timesteps_per_batch (the number of timesteps to run per batch (horizon)):* 0.3

*used architecture for policy representation*: Multi Layer Perceptron - 2 layers of 64

*n_episodes (Number of trajectories (episodes) to record):* 10

*traj_limitation (the number of trajectory to use):* 7000

*Lambda (the weight for the entropy loss):* 1

*total_timesteps (The total number of samples to train on):* 100000

Training and evaluation of the models in this project has been done using Colab virtual machines by google. Training time in average took the most for Generative Adversarial Imitation Learning method with an average of 56 minutes. Second place goes to Soft Actor Critic method with an average of 29 minutes and finally Proximal Policy Optimization method took about 14 minutes on average to be trained for these datasets.

5. *Evaluation and Strategy Creation*

To evaluate each model's performance in this project we use *total reward value* and *total cost value* of the strategy execution. By adding total reward to starting value of the account and subtracting total cost from it, we get final value of the account. To make the policies more interpretable, we have added two figures to our results which denote the buying and selling points of the agent with green and red dots respectively on the price movements and account balance charts. Note that as we have gathered our data from Binance, we have taken into account its current transaction fee per trade (0.75 percent of the trade value) when simulating the market.

Strategy Creation procedure is pretty straightforward. We took 60 rows of records from now and past data (which make our agent's state) and decide based on them and the trained policy the next action. After 4 hours the system retries this operation to decide for the next coming 4 hours. It will accumulate positive or negative rewards and also the costs of the actions through the test span. Notice that our position size is variable in execution phase and it is determined using our policy. This is one of the advantages that we get by using continues space reinforcement learning. At the final step of the strategy we sell whatever we have. In the next section we look at the results of our experiments with our models.

## V. EXPERIMENTAL RESULTS

Here we look at three different cryptocurrency markets that we have studied, separately. This section has three subsections relative to each cryptocurrency pair. At each subsection, first, we have a graph of the pair's price movements through the time span that we scrutinize it. Then, we report our results from running each algorithm in a table. After each table, we have two graphs, the first one shows the pair's price movements through its test phase with the buying and selling actions denoted on it. Second one shows account balance through test time span. We have denoted buying and selling actions on the second graphs too. Green dots represent buying actions and red dots represent selling actions.

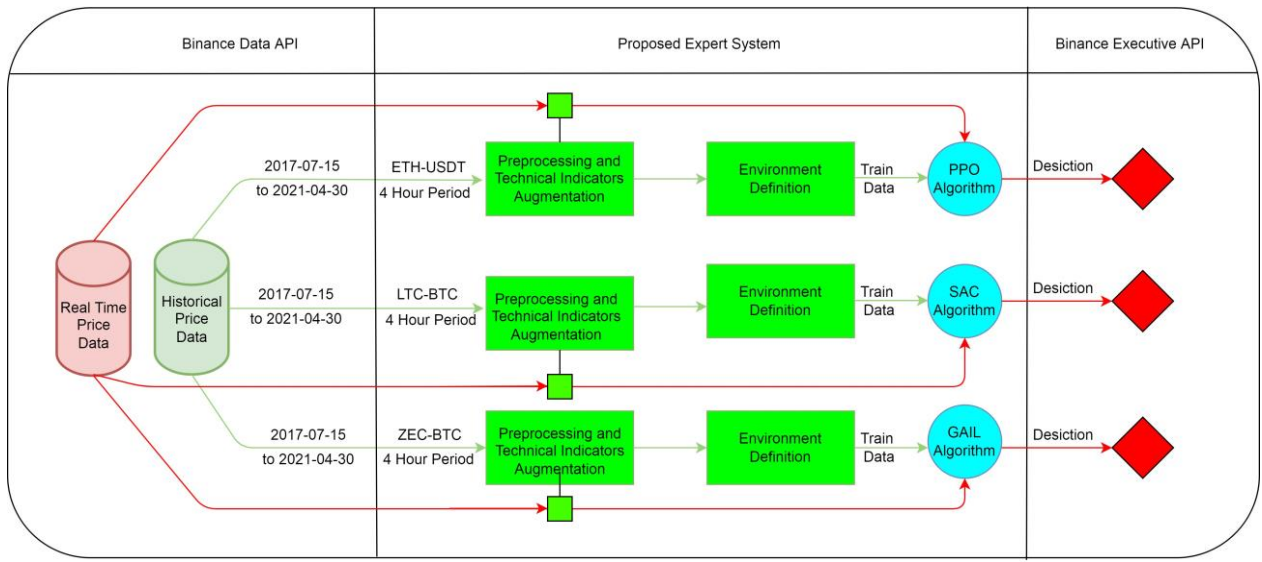

Figure 5. Overall Structure of The Proposed Expert System for DRL Methods.

Green lines indicate train phase and red lines indicate exertion phase

A. *ETH-USDT:*

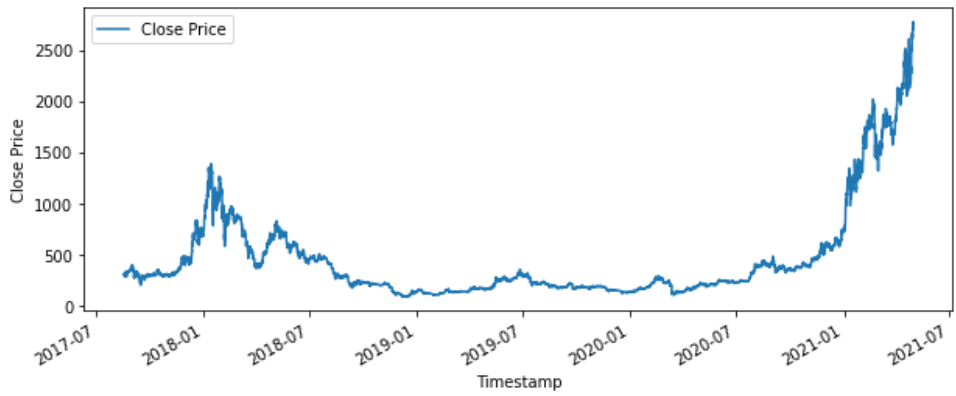

Figure 6. Close Price for ETH-USDT from 2017-07 to 2021-05

| Begin Account Value | 10000 |
| --- | --- |
| End Account Value | 14546.08504638672 |
| Total Cost | 649.9545043945312 |
| Total Trades | 474 |
| Start Date/End Date | 2021-02-24/2021-05-01 (66 Days) |

Table 1. Information Regarding PPO Test on ETH-USDT

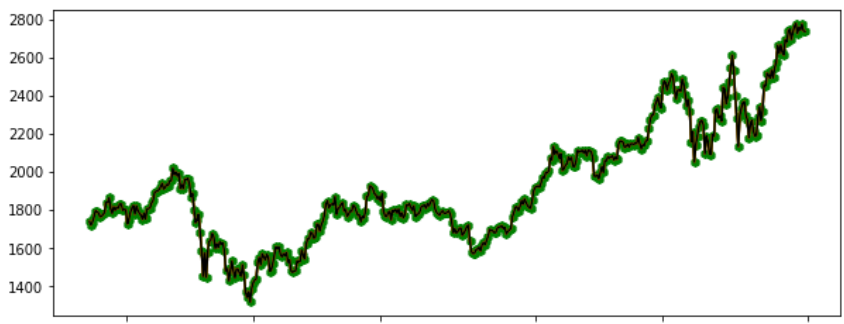

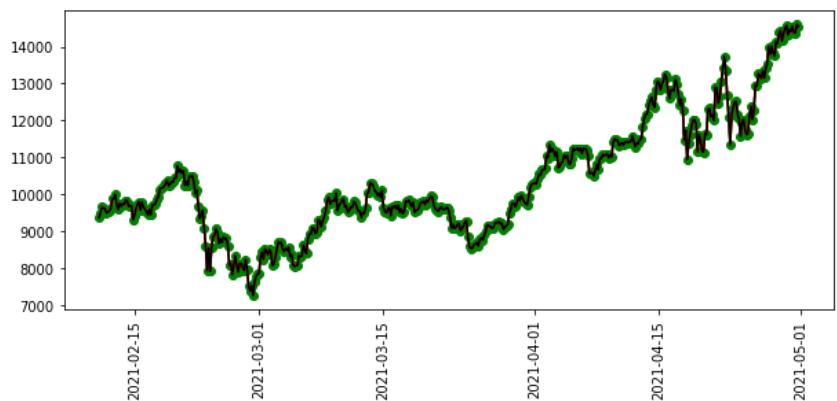

Figure 7. Close Price for ETH-USDT in Test Data
Figure 8. Account Balance in PPO Algorithm for ETH-USDT in Test Data

| Begin Account Value | 10000 |
| --- | --- |
| End Account Value | 10000.0 |
| Total Cost | 0 |
| Total Trades | 0 |
| Start Date/End Date | 2021-02-24/2021-05-01 (66 Days) |

Table 2. Information Regarding SAC Test on ETH-USDT

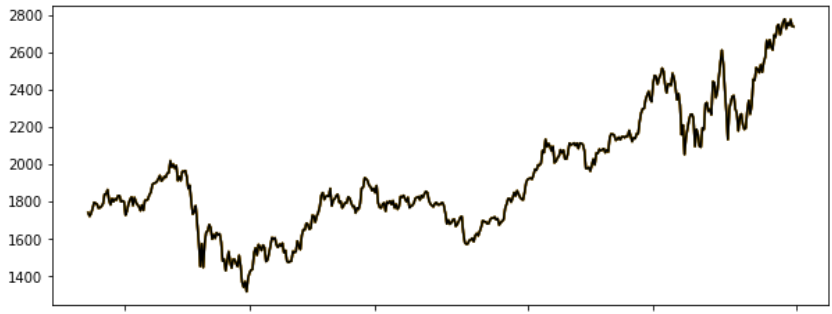

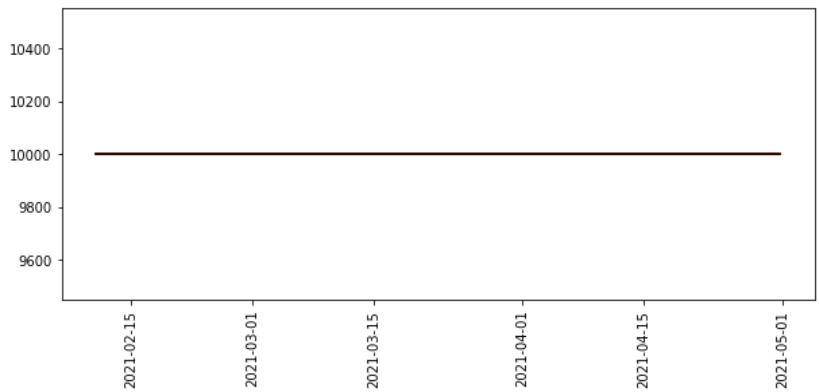

Figure 9. Close Price for ETH-USDT in Test Data
Figure 10. Account Balance in SAC Algorithm for ETH-USDT in Test Data

| Begin Account Value | 10000 |
| --- | --- |
| End Account Value | 14171.616086510576 |
| Total Cost | 632.2463374660161 |
| Total Trades | 472 |
| Start Date/End Date | 2021-02-24/2021-05-01 (66 Days) |

Table 3. Information Regarding GAIL Test on ETH-USDT

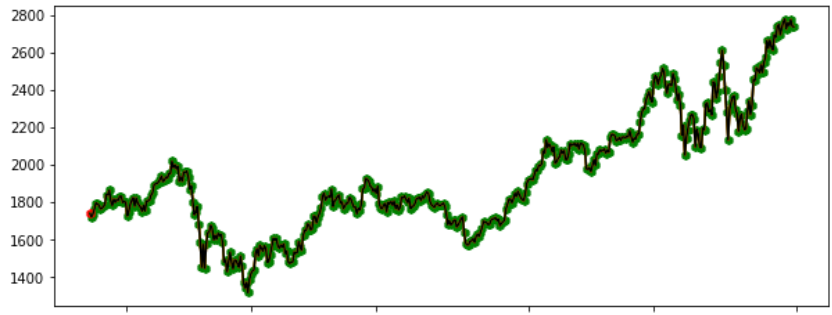

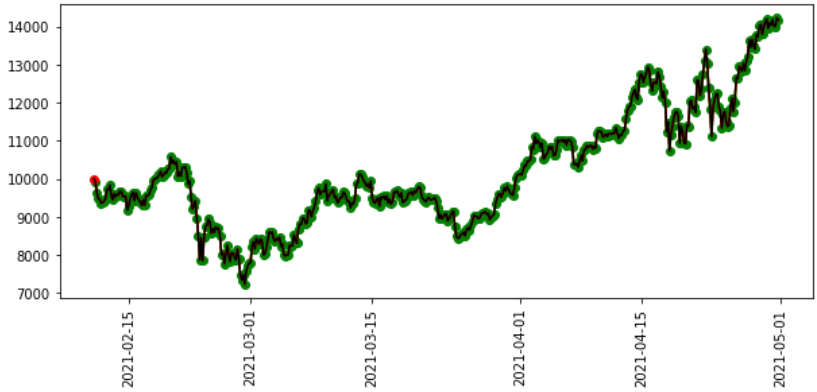

Figure 11. Close Price for ETH-USDT in Test Data
Figure 12. Account Balance in GAIL Algorithm for ETH-USDT in Test Data

## B. LTC-BTC

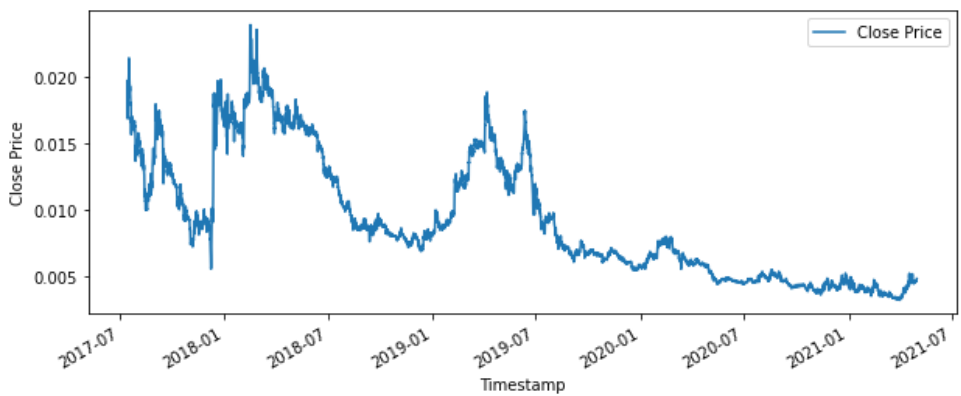

Figure 13. Close Price for LTC-BTC from 2017-07 to 2021-05

| Begin Account Value | 1.0 |
|---|---|
| End Account Value | 1.1792819791357025 |
| Total Cost | 0.09588345007505264 |
| Total Trades | 364 |
| Start Date/End Date | 2021-02-24/2021-05-01 (66 Days) |

Table 4. Information Regarding PPO Test on LTC-BTC

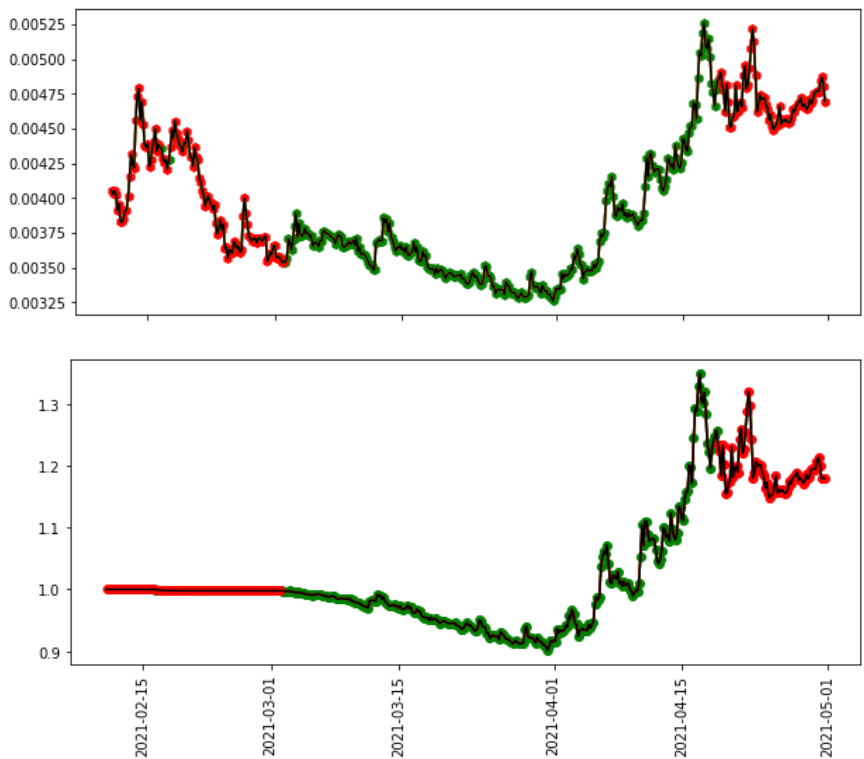

Figure 14. Close Price for LTC-BTC in Test Data
Figure 15. Account Balance in PPO Algorithm for LTC-BTC in Test Data

| Begin Account Value | 1.0 |
| --- | --- |
| End Account Value | 1.1144074097550005 |
| Total Cost | 0.08855678078523896 |
| Total Trades | 473 |
| Start Date/End Date | 2021-02-24/2021-05-01 (66 Days) |

Table 5. Information Regarding SAC Test on LTC-BTC

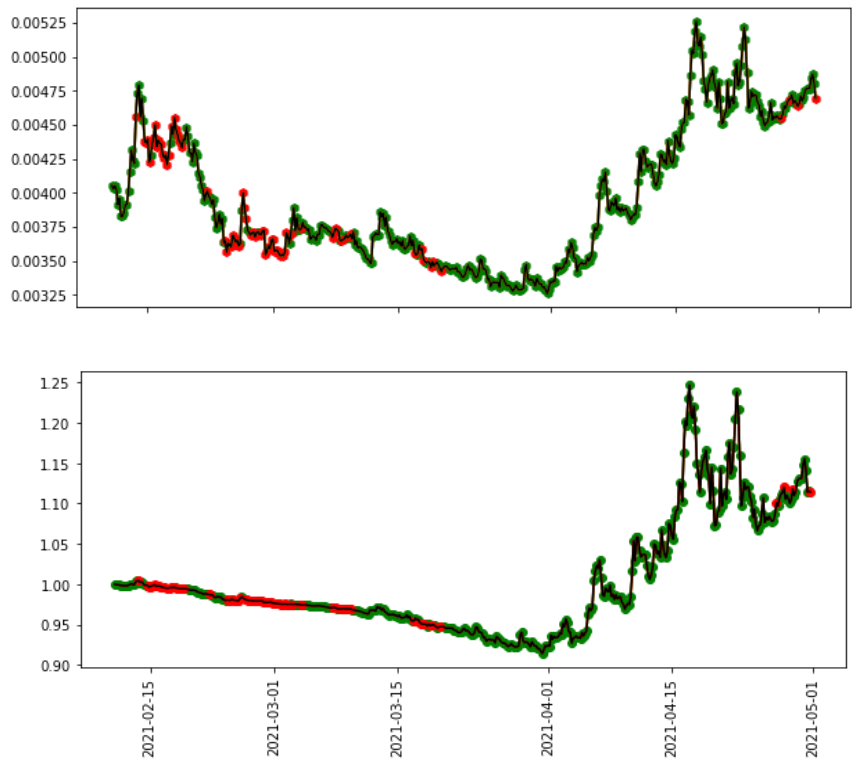

Figure 16. Close Price for LTC-BTC in Test Data
Figure 17. Account Balance in SAC Algorithm for LTC-BTC in Test Data

| Begin Account Value | 1.0 |
| End Account Value | 0.9174084284001103 |
| Total Cost | 0.07326343385358162 |
| Total Trades | 370 |
| Start Date/End Date | 2021-02-24/2021-05-01 (66 Days) |

Table 6. Information Regarding GAIL Test on LTC-BTC

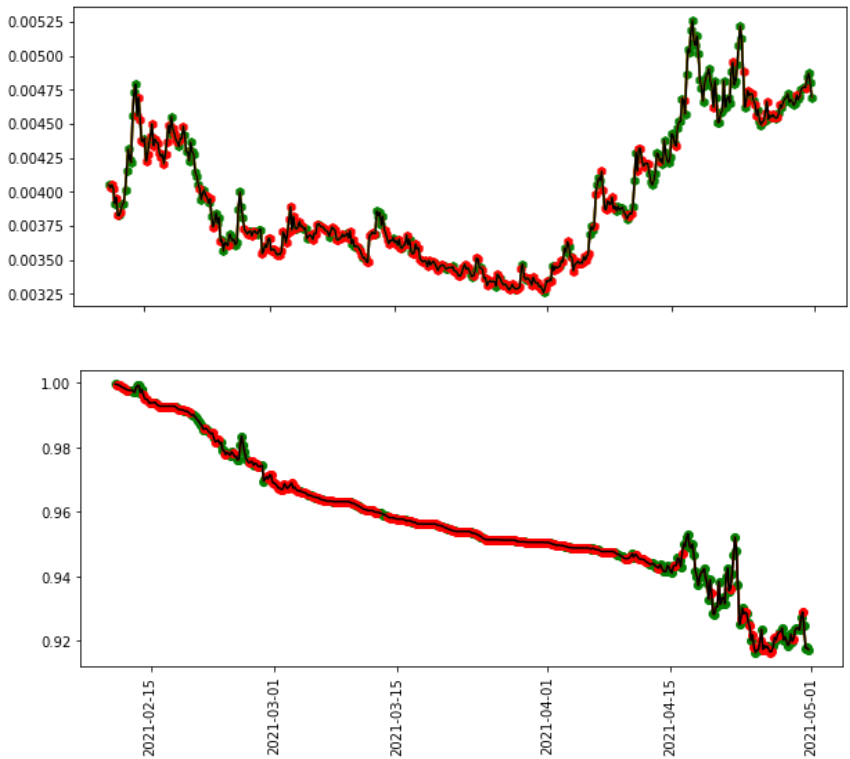

Figure 18. Close Price for LTC-BTC in Test Data
Figure 19. Account Balance in GAIL Algorithm for LTC-BTC in Test Data

## C. ZEC-BTC

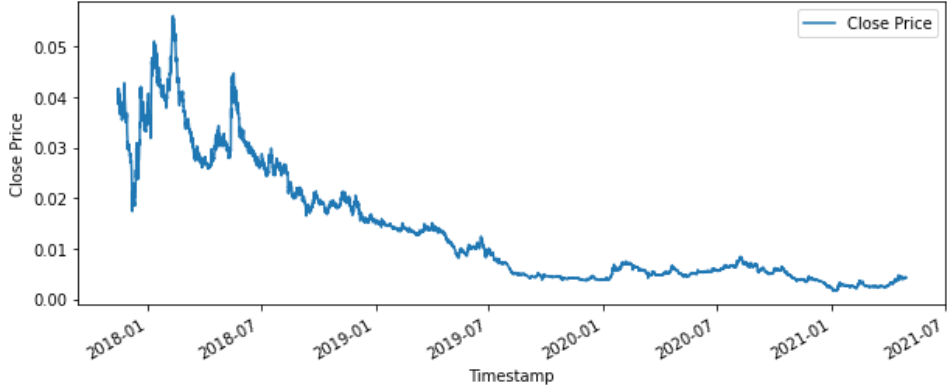

Figure 20. Close Price for ZEC-BTC from 2017-07 to 2021-05

| Begin Account Value | 1.0 |
| --- | --- |
| End Account Value | 1.071781530917556 |
| Total Cost | 0.027391680507381718 |
| Total Trades | 474 |
| Start Date/End Date | 2021-02-24/2021-05-01 (66 Days) |

Table 7. Information Regarding PPO Test on ZEC-BTC

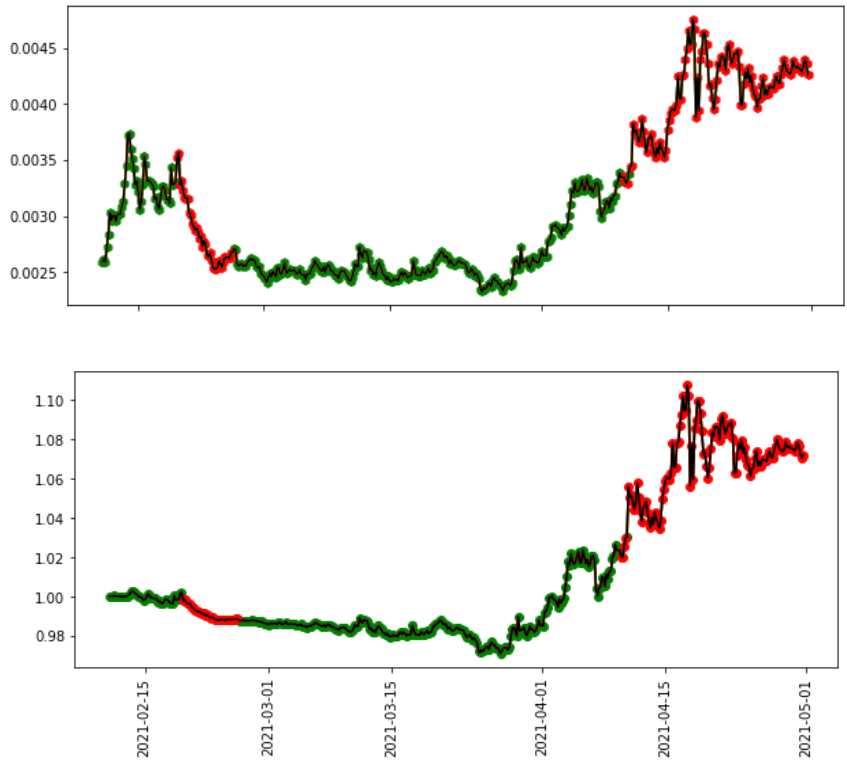

Figure 21. Close Price for ZEC-BTC in Test Data
Figure 22. Account Balance in PPO Algorithm for ZEC-BTC in Test Data

| Begin Account Value | 1.0 |
| End Account Value | 1.0954109611514145 |
| Total Cost | 0.026081938917657897 |
| Total Trades | 474 |
| Start Date/End Date | 2021-02-24/2021-05-01 (66 Days) |

Table 8. Information Regarding SAC Test on ZEC-BTC

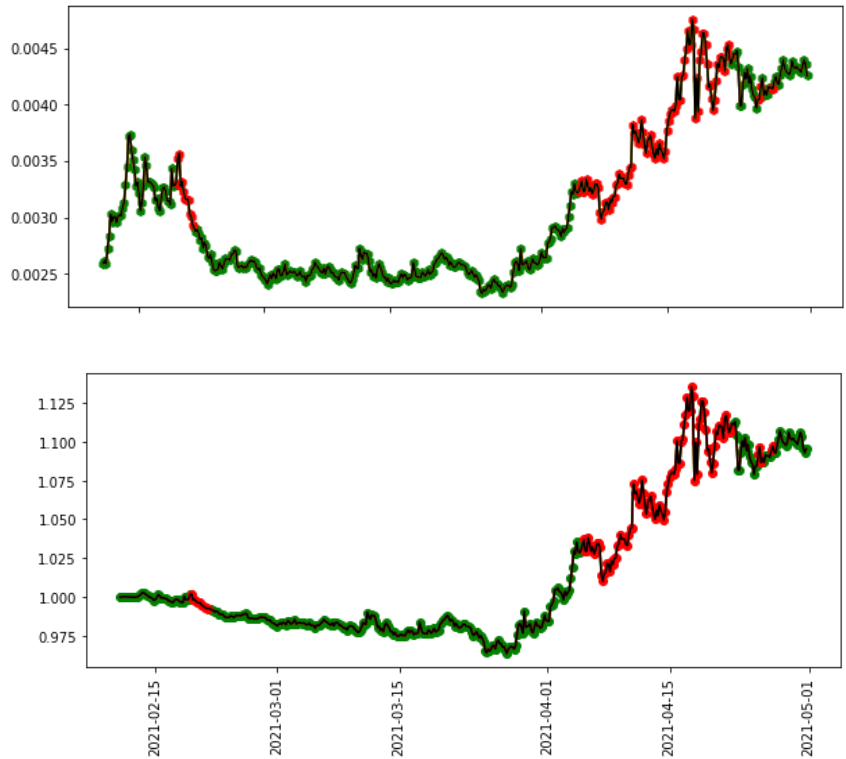

Figure 23. Close Price for ZEC-BTC in Test Data
Figure 24. Account Balance in SAC Algorithm for ZEC-BTC in Test Data

| Begin Account Value | 1.0 |
| End Account Value | 0.980018116607434 |
| Total Cost | 0.021079767410144522 |
| Total Trades | 433 |
| Start Date/End Date | 2021-02-24/2021-05-01 (66 Days) |

Table 9. Information Regarding GAIL Test on ZEC-BTC

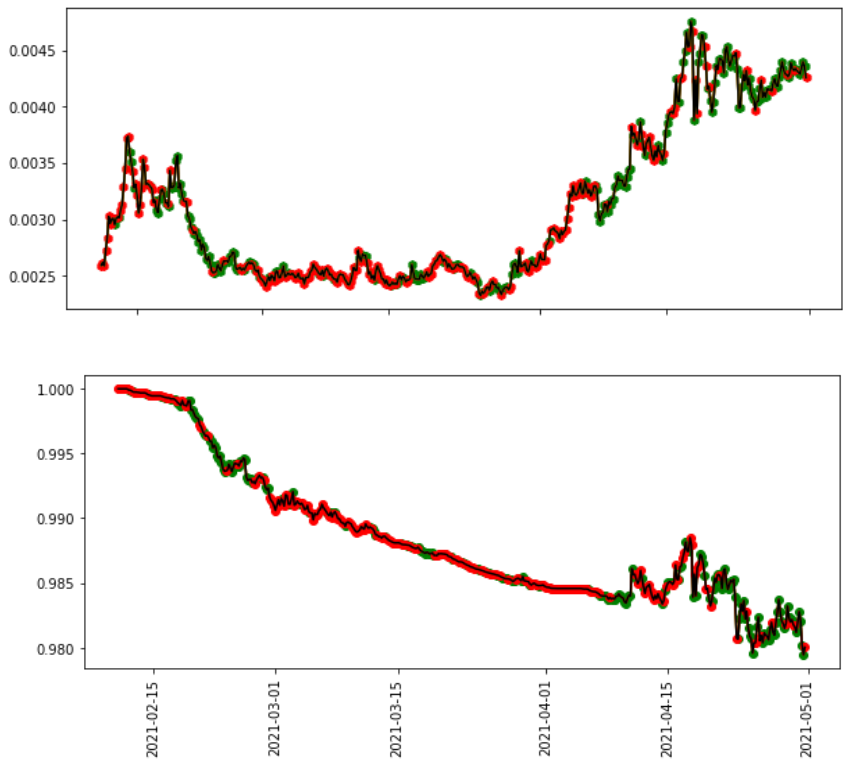

Figure 25. Close Price for ZEC-BTC in Test Data
Figure 26. Account Balance in GAIL Algorithm for ZEC-BTC in Test Data

## VI. DISCUSSION

In this section we discuss how our models have been performing through different market conditions. We point to a hyperparameter which can be used to implicitly adjust the involved risk in designed strategies. We also discuss the limitations and interpretability issue of this AI-based approach and point to how this system can be used to exploit the market.

All our three studied cryptocurrency pairs show different market conditions through our models train phases (i.e. all of them have bullish, bearish and range movements), although frequency of these moves are not equal and this could make effects on the results. By comparing figure 14 and figure 15 we can observe how PPO algorithm has avoided the bearish market in early weeks of the test data. This observation can also be made with comparing figure 23 with figure 24. In ETH-USDT data we can see our models have not been performing well. PPO algorithm has classified the whole time span as bullish market and has signalled buying all the time. SAC algorithm too has not entered any trades in this time span, hence reporting no profit or loss. GAIL algorithm for ETH-USDT has been performed better than other pairs (maybe because of simplicity of the policy in the test data conditions). Another interesting observation that can be made with figures 14 and 15 and also with figures 16 and 17 is how models have been buying the asset before its rise in early April 2021.

An important aspect of the introduced models is their hyperparameter tuning. Some specific methods that try to minimize the number of these hyperparameters have been used in this research work, but they still exist. A proper approach for tuning these variables could be using grid search or metaheuristics. An interesting usage of one of these hyperparameters is to adjust the risk involved with the designed strategy by means of it. This hyperparameter is Maximum Buying Amount. It gets multiplied by action space each time the agent generates a new action. By setting it in low values we can make the agent's position size smaller in each trade, hence decreasing the risk and by setting it in high values we can encourage the agent to take higher risks. This value in our experiments has been set in a way that it could assign a position size equal to its initial balance at its maximum.

An important issue in practical usage of this system is its interpretability. Interpretability of artificial intelligence is a rising concern in new applications of it. In this research work we tried to demonstrate agents actions in test phase, but still the investor may not trust the policies derived from "black box" models.

Another interesting approach that has been studied in this work is the potential of using IRL and GAN to imitate expert's behaviour. We haven't used GAIL algorithm with human expert's behaviour, but a comprehensive study of this method is suggested for future study of extracting knowledge from expert generated trajectories.

All of what has been discussed till now are theoretical arguments, implications of these models look very attractive but it definitely will bring up new issues and more research needs to be done. Many exchanges nowadays allow automatic traders to act in their provided markets. One can use these exchanges data and process them inside the introduced schemes and decides and trades based on them in hope of profit. As cryptocurrency markets are almost always available (i.e. 24/7) using a dedicated server can find trade opportunities and acts on them automatically.

## VII. CONCLUSIONS AND FUTURE WORKS

The impact of artificial intelligence's applications in many areas are promising for a more efficient and prosperous future. In this study we looked at three different deep reinforcement learning approaches to help investors to make their decisions in some new emerging international markets in a more data driven and autonomous manner. PPO and SAC models showed positive returns in unseen data. Our maximum profit factor was 1.45 for ETH-USDT by PPO in 66 days. Although GAIL models didn't showed positive return in this study but it has to be take into account we have imitated non-expert trajectories in this research work. We have showed how these models can be used with behaviours of successful traders in order to capture their strategy. It's obvious more research needs to be done in this area. Most of resulting strategies in this research work still lack "smoothness" in their final balance graphs and hence showing large potential risks to be implemented. Designing a full autonomous trading system surely involves more concerns than the ones we had simplified in this research work, like market liquidity issues.

In this research work we have provided a data scientific approach for the strategy design procedure for cryptocurrencies which yields positive returns in different market conditions by using only price and volume as input. We also presented a framework for using inverse reinforcement learning to capture intuitional techniques which human experts use, for application in the financial markets.

As we can see, there seems a predictability potential in a highly complex system like financial markets by means of deep reinforcement learning. For future works, our suggestions include:

1. Combining fundamental information with technicals to improve the accuracy
2. Ensembling different approaches in machine learning to decrease the bias of the whole system
3. Using social networks data streams to obtain an accumulated view on public opinion on different assets
4. Using Deep neural networks to feature extraction from raw data
5. Using machine learning approaches for risk management in a collateral system to decision making
6. Doing a comprehensive study of inverse reinforcement learning for expert knowledge extraction in this field and of generative adversarial networks for generating similar trajectories for training

Besides what we have discussed about financial markets, it seems deep reinforcement learning models can be used in many other chaotic natured problems which share some of their data characteristics with financial data. These fields could include supply chain support, Business affairs with public opinions, public views on political issues and many other use cases.

# Appendix A: Used Technical Indicators and Their Formulas

In this appendix we introduce the technical indicators used in this project and their respective formulas.

### Commodity Channel Index (CCI):

$$CCI = \frac{Typical\ Price - MA}{0.015 * Mean\ Deviation}$$

where:

$$Typical\ Price = \sum_{i=1}^{P}((High + Low + Close)/3)$$
$$P = Number\ of\ Periods$$
$$MA = Moving\ Average$$
$$Moving\ Average = (\sum_{i=1}^{P} Typical\ Price)/P$$
$$Mean\ Deviation = (\sum_{i=1}^{P} |Typical\ Price - MA|)/P$$

We have used this indicator in 14 and 30 periods in this project.

### Relative Strength Index (RSI):

$$RSI_{Step\ One} = 100 - \left[\frac{100}{1 + \frac{Average\ Gain}{Average\ Loss}}\right]$$

The average gain or loss used in the calculation is the average percentage gain or loss during a look-back period. The formula uses a positive value for the average loss.

Once there is first step data available, the second part of the RSI formula can be calculated. The second step of the calculation smooths the results:

$$RSI_{Step\ Two} = 100 - \left[\frac{100}{1 + \frac{(Previous\ Average\ Gain * (Period - 1)) + Current\ Gain}{-((Previous\ Average\ Loss * (Period - 1)) + Current\ Loss)}}\right]$$

We have used this indicator in 14 and 30 periods in this project.

### Directional Movement Index (DMI):

$$DX = \left(\frac{|DI^+ - DI^-|}{|DI^+ + DI^-|}\right) * 100$$

where:

$$DI^+ = \left(\frac{Smoothed\ (DM^+)}{ATR}\right) * 100$$
$$DI^- = \left(\frac{Smoothed\ (DM^-)}{ATR}\right) * 100$$
$$DM^+ (Directional\ Movement) = Current\ High - Previous\ High$$
$$DM^- (Directional\ Movement) = Previous\ Low - Current\ Low$$
$$ATR = Average\ True\ Range$$
$$Smoothed\ (x) = \sum_{t=1}^{Period} x - \frac{\sum_{t=1}^{Period} x}{Period} + CDM$$
$$CDM = Current\ DM$$

We have used this indicator with period=14 in this project.

### Moving Average Convergence Divergence (MACD):

$$MACD = EMA_{12\ Period} - EMA_{26\ Period}$$

### Bollinger Band®:

$$Boll = \frac{Boll_U + Boll_D}{2}$$
$$Boll_U = MA(TP, n) + m * \sigma[TP, n]$$
$$Boll_D = MA(TP, n) - m * \sigma[TP, n]$$

where:

$$Boll_U = Upper\ Bollinger\ Band$$
$$Boll_D = Lower\ Bollinger\ Band$$
$$MA = Moving\ Average$$
$$TP\ (Typical\ Price) = (High + Low + Close)/3$$
$$n = Number\ of\ Days\ in\ Smoothing\ Period\ (Typically\ 20)$$
$$m = Number\ of\ Standard\ Deviations\ (Typically\ 2)$$
$$\sigma[TP, n] = Standard\ Deviation\ over\ Last\ n\ Periods\ of\ TP$$